\documentclass[runningheads]{llncs}
\usepackage{tabularx}
 \usepackage{array,multirow,graphicx}
\usepackage{graphicx}
\usepackage[noend]{algpseudocode}
\usepackage{algorithm,algorithmicx}

\algrenewcommand\algorithmicrequire{\textbf{Input:}}

\newcolumntype{P}[1]{>{\centering\arraybackslash}p{#1}}
\usepackage{xcolor}
\usepackage{amsmath}

\usepackage{hyperref}

\begin{document}
%
\title{QtSeg: an annotation-free method for state-of-the-art segmentation in histopathological images}
\title{Pan-cancer whole slide image segmentation by leveraging novel weak annotations}
\title{QtSeg: a weakly supervised whole slide image pan-cancer segmentation tool}
\title{Weakly supervised pan-cancer segmentation tool}
\titlerunning{QtSeg: boosting weakly supervised segmentation}
\titlerunning{Pan-cancer WSI tumor segmentation with novel weak annotations}
\titlerunning{QtSeg: a weakly supervised WSI pan-cancer segmentation tool}
\titlerunning{Weakly supervised pan-cancer segmentation tool}
%
%
\author{
Marvin Lerousseau\inst{1,2} \and
Marion Classe\inst{2,3} \and
Enzo Battistella\inst{1,2} \and
Th\'eo Estienne\inst{1,2} \and
Th\'eophraste Henry\inst{1,2} \and
Amaury Leroy\inst{1,2} \and
Roger Sun\inst{1,2} \and
Maria Vakalopoulou\inst{1,2} \and
Jean-Yves Scoazec\inst{3} \and
Eric Deutsch\inst{1,2} \and
Nikos Paragios\inst{4}
}
%
\authorrunning{M. Lerousseau et al.}
%

%

\institute{
Paris-Saclay University, CentraleSup\'elec, 91190, Gif-sur-Yvette, France \and
Paris-Saclay University, Gustave Roussy, Inserm, 94800, Villejuif, France \and
Paris-Saclay University, Gustave Roussy, Pathology department, 94800, Villejuif, France \and
TheraPanacea, 75014, Paris, France
}
\maketitle              
\begin{abstract}
The vast majority of semantic segmentation approaches rely on pixel-level annotations that are tedious and time consuming to obtain and suffer from significant inter and intra-expert variability. 
To address these issues, recent approaches have leveraged categorical annotations at the slide-level, that in general suffer from robustness and generalization.
In this paper, we propose a novel weakly supervised multi-instance learning approach that deciphers quantitative slide-level annotations which are fast to obtain and regularly present in clinical routine. 
The extreme potentials of the proposed approach are demonstrated for tumor segmentation of solid cancer subtypes.
The proposed approach achieves superior performance in out-of-distribution, out-of-location, and out-of-domain testing sets.

\end{abstract}
\section{Introduction}
Semantic segmentation relies on supervised learning where images are paired with manually curated maps~\cite{garcia2017review}. 
This has also been the case for medical imaging~\cite{litjens2017survey}, and computational pathology~\cite{li2020deep}.
In digital pathology, despite efforts to annotate and provide ground-truth segmentations for some cancer types, the lack of fine-grained annotations hinders the development of automatic segmentation tools for solid cancers. 
Unfortunately, pathologists-curated datasets are fairly small, given that the biggest open-source dataset with pixelwise annotations contains 400 whole slide images~\cite{litjens20181399}. 
Consequently, the generalization performance of the learnt models is limited and they fail to grasp the underlying clinical heterogeneity regarding tissue preparation, slide digitization, and tissue structures~\cite{campanella2019clinical}.
The lack of pixelwise annotations could be explained by the tedious and cumbersome facet of the task, aggravated by the lack of availability of pathologists~\cite{metter2019trends}.

Despite this expertise shortage, we observe an increasing availability of whole slide images~\cite{stathonikos2013going,zarella2019practical}. 
In recent years, more and more efforts are devoted towards publicly available data.
The Cancer Genome Atlas (TCGA)~\cite{tomczak2015cancer} which counts 30072 whole slide images from 32 cancer subtypes is the most prominent example.
Whole slide images lack in general annotations but are routinely associated with weak clinical variables such as labels at the slide-level. Slide labels are abundant weak annotations for WSIs.

There exists a rich literature on segmentation models trained from using categorical labels at the slide-level~\cite{hou2016patch,coudray2018classification,oquab2015object,schmauch2020deep,kim2019deep,campanella2019clinical,campanella2018terabyte,lerousseau2020weakly}.
Most of them rely on multiple instance learning (MIL)~\cite{dietterich1997solving,foulds2010review} where an embedder maps instances in an embedding space, which are further processed by a pooling operator.

In this work, we exploit eyeball estimates of percentages of tumor in whole slide images for training tumor segmentation models. 
Such quantities are routinely obtained during tumor purity estimation, a process ensuring that a tissue sample has enough neoplastic material for further genetic tests. 
We harness such labels through a highly modular algorithm derived from MIL to train segmentation models (\autoref{sec:methods}). Large-scale experiments involving almost all types of solid primary tumors are performed on entirely public data, including the percentages estimates (\autoref{sec:experiments}).

\section{Methods}
\label{sec:methods}
Let us consider a set of training whole slide images $(x_k)$.
Specifically, each whole slide image $x_k$ is denoted by its constituents $(x_{k, i})$, such as pixels or tiles.
Our goal is to learn a $\theta$-parametrized model $f_\theta$ that maps any $x_k$ to its (unknown) tumor segmentation map $\hat{y}_k$, apparent to a coarse map $(\hat{y}_{k, i})_i$. 
Each training sample $x_k$ has a label $p_k\in[0, 100]\%$ representing the percentage of tumor within $x_k$. 
We denote by $\max\limits_{p_k}\{f_\theta(x_k)\}$ the $p_k\%$ maximal values of the predicted coarse map $\{f_\theta(x_{k, i}); x_{k, i} \in x_k\}$ and by $\min\limits_{100-p_k}\{f_\theta(x_k)\}$ the $100-p_k\%$ minimal values.

We now present \textsc{WeSeg} (\textbf{W}eakly sup\textbf{E}rvised \textbf{Seg}mentation), an algorithm whose goal is to train $f_\theta$ by ensuring that the percentage of predicted tumor is equal to the percentage annotation.
To do so, we adopt a recursive training mechanism for each sample $x_k$.
Specifically, \textsc{WeSeg} builds a proxy ground-truth map that contains $p_k\%$ of tumor pixels by assigning a value of 1 to the $p_k\%$ of pixels with maximum predicted probabilities $(f_\theta(x_{k, i}))_i$, otherwise 0.

Equivalently, given an elementwise loss function $L$ such as binary cross-entropy, \textsc{WeSeg} aims at minimizing the following empirical risk for a sample $x_k$ with percentage annotation $p_k$:
$$
e = \sum_{i \in \max\limits_{p_k}\{f_\theta(x_k)\}} L\left(f_\theta(x_{k, i}), 1\right) + \sum_{i \in \min\limits_{100-p_k}\{f_\theta(x_k)\}} L\left(f_\theta(x_{k, i}), 0\right)
$$
By construction, this methodology ensures that an error signal of $0$ for a sample $(x_k, p_k)$ implies that $p_k\%$ of output probabilities are of value 1, and $100-p_k\%$ of value 0, which is precisely the meaning of the percentage annotation. 
After training, the learned segmentation model can be inferred on new WSIs without any percentage annotation.
Besides, while our notation uses percentage annotations, this approach can be used with count-based weak annotations.


As illustrated by the public annotations used in our experiments, the percentage annotations can be noisy. This is notably a consequence of the difficulty of estimating a surface to the human eye.
To take this uncertainty into account, we introduce four parameters $r_{low}$, $r_{high}$, $a_{low}$ and $a_{high}$ such that for a sample with percentage annotation $p_k$:
\begin{itemize}
	\item probabilities below the $((1 - r_{low}) * p_k - a_{low})$th percentile are assigned 0
	\item probabilities above the $((1 + r_{high}) * p_k + a_{high})$th percentile are assigned 1
	\item remaining probabilities are discarded from the error signal computation
\end{itemize}
With $r_{low}=r_{high}=a_{low}=a_{high}=0$ this formulation is equivalent to the original version above. Generally speaking, the $r$ (resp. $a$) parameters control a relative (resp. absolute) margin around the annotation. In practice, these values are selected either empirically or by finding bias in annotations.

Our approach refers to a modular, scalable and network architecture-free framework. 
Its main strength lies in the ability to progressively construct ground truth from very weakly annotated data, and recursively update the associated network architecture towards better and better performance. 
The underlying principle employs the concept Highest Confidence First where progressively throughout this recursive process, data samples are labeled with accurate labels and fed back to the network for end-to-end retraining. 
For demonstration purposes we have adopted a conventional ResNet~\cite{he2016identity} architecture, that by no means restricts the nature of architectures that can be accommodated such as Unet~\cite{ronneberger2015u}, GAN, etc from the proposed versatile paradigm.


\section{Experiments}
\label{sec:experiments}
Large-scale experiments are conducted for pan-cancer tumor segmentation in WSIs.
Several weakly supervised learning frameworks are benchmarked (\autoref{sec:benchmarks})
by optimizing the same architecture with common-hyperparameters (\autoref{sec:archi_details}) on a shared training set (\autoref{sec:datasets}). Results are reported (\autoref{sec:results}) for multiple testing sets (\autoref{sec:datasets}).

\subsection{Benchmarked approaches}
\label{sec:benchmarks}
\begin{itemize}
	\item \textbf{AttentionMIL}~\cite{ilse2018attention} classifies WSIs with an attention module. As suggested by the authors, for a given set of tiles with computed attention weights $(a_k)$, the scores are obtained by min-max scaling the weights.
Additionally, if a WSI is predicted as normal, all tile predictions are set to 0.
The attention module was a two layer feedforward neural network with 128 hidden neurons.

	\item \textbf{AlphaBetaMIL}~\cite{lerousseau2020weakly} learns an instance-level MIL model with two parameters $\alpha$ and $\beta$ which are fixed and common for all training samples.
For WSIs with label 0, their approach assigns all pixels to 0. Otherwise, $\alpha$\% of pixels with highest probability are assigned 1, while the $\beta$\% lowest are assigned 0.
	10 configurations were uniformly sampled for parameters $\alpha$ and $\beta$.
	
	\item A \textbf{supervised} approach is emulated by considering all slides with annotation of either $0\%$ or $100\%$ and discarding others. With a $0\%$ annotation, the ground-truth map can be inferred as filled with benign tissue. On the contrary, with an annotation of $100\%$, the ground-truth can be inferred as filled with $1$. This approach is therefore apparent to a supervised learning one with pixelwise ground-truth segmentation maps. 
	
	\item The \textbf{proposed} approach is trained with all percentages annotations. The values of $r_{low}$, $r_{high}$, $a_{low}$ and $a_{high}$ were set to $0$.
\end{itemize}

\subsection{Datasets}
\label{sec:datasets}
\subsubsection{Training data}
The training set was built using the snap-frozen whole slide images TCGA data referring to the 32 most common solid cancer subtypes with 3 of them being excluded from training and used as testing set, leading to a training set of 18306 WSIs from 10903 patients.
The distribution of the number of whole slide images and patients per subtype can be found at the supplementary Table 1.
The 10903 patients were further split into 70\%, 10\%, and 20\% cases for training, validation, and test sets respectively.

Two types of annotations were retrieved. 
Firstly, binary labels indicating whether each WSI is of normal type (0), \textit{i.e.} with no apparent neoplastic tissue, otherwise 1. 2248 out of the 12783 WSIs are found normal.
Secondly, we retrieved publicly available annotations of percentages of tumor for all but 235 out of the 18306 WSIs. These annotations were available in the TCGA repository and identified by "\textit{percent\_tumor\_cells}".

\subsubsection{Testing performance}
In order to derive a meaningful assessment of the generalization capabilities of the benchmarked models, several testing sets were constituted. 
First, 175 WSIs were sampled from the hold-out testing set of the training cohort, \textit{i.e.} 5 WSIs from each of the 29 training subtypes.
2 pathologists curated ground-truth tumor maps with exhaustive annotations.
This set is identified as the out-of-distribution testing set.

Similarly, 24 pixelwise WSI annotations were curated on the 3 cancer subtypes excluded from the training cohort, \textit{i.e.} cervix squamous cell, rectum adenocarcinoma, and glioblastoma multiforme.
Both rectum and cervix locations are unseen in the training set. 
Besides, the training set contains only low grade glioblastoma for the brain location. 
Consequently, these 3 testing cohorts provide insight about the out-of-location performance.

Finally, out-of-domain performance was assessed on Formalin-fixed paraffin-embedded (FFPE) slides from 2 open-source datasets with public pixelwise ground-truth: DigestPath for colon adenocarcinoma~\cite{li2019signet}, and PAIP2019 for liver hepatocellular carcinoma~\cite{kim2020paip}. 
FFPE tissue displays dissimilarity with frozen tissue, therefore providing insight about out-of-domain generalization.

\subsubsection{WSI pre-processing}
All WSI were tiled at a magnification of 10x into 512 pixel-wide tiles with 128 overlapping pixels on each side using~\cite{martinez2007libvips}. Tiles detected as background were completely discarded from the study (including performance measures), \textit{i.e.} those where at least 90\% of pixels have both red, green, and blue channels above 200.

\subsection{Optimization details}
\label{sec:archi_details}
All approaches are used to train a ResNet50~\cite{he2016deep} architecture pretrained on Imagenet~\cite{deng2009imagenet}.
At each training step, 30 tiles were randomly sampled for each of 8 whole slide images, for a total batch size of 240.
Each tile was data-augmented with random vertical and horizontal flips, color jitter with brightness $0.1$, contrast $0.1$, saturation $0.1$, and hue $0.01$, standard scaled with mean and variance computed from the training set, and random cropped into 224 pixel-wide tiles.
In order to produce segmentation of higher granularity, the penultimate layer of ResNet50 was discarded, \textit{i.e.} global adaptive pooling. For a 224 pixel-width input, the output features maps are thus of size $7\times7$. A final linear classification layer was applied elementwise, producing outputs of size $7\times7$ instead of $1$.
Early stopping was triggered when no validation loss improvement was found for 50 epochs.
Error signals were computed with binary cross-entropy and weights were updated with the Adam optimizer~\cite{kingma2014adam}.
An independent random search was performed for all learning algorithms for learning rate selection within a range of $[1e^{-6}, 1e^{-3}]$ in a log-uniform fashion. Final learning rates were selected with the best achieved validation losses.
Each random search was given a budget of 5 days on 2 V100. The total training time of our experiments is 117 V100$\cdot$days.
All experiments was performed with pytorch~\cite{paszke2019pytorch} v1.7.1.

\begin{figure}[!ht]
\includegraphics[width=\textwidth]{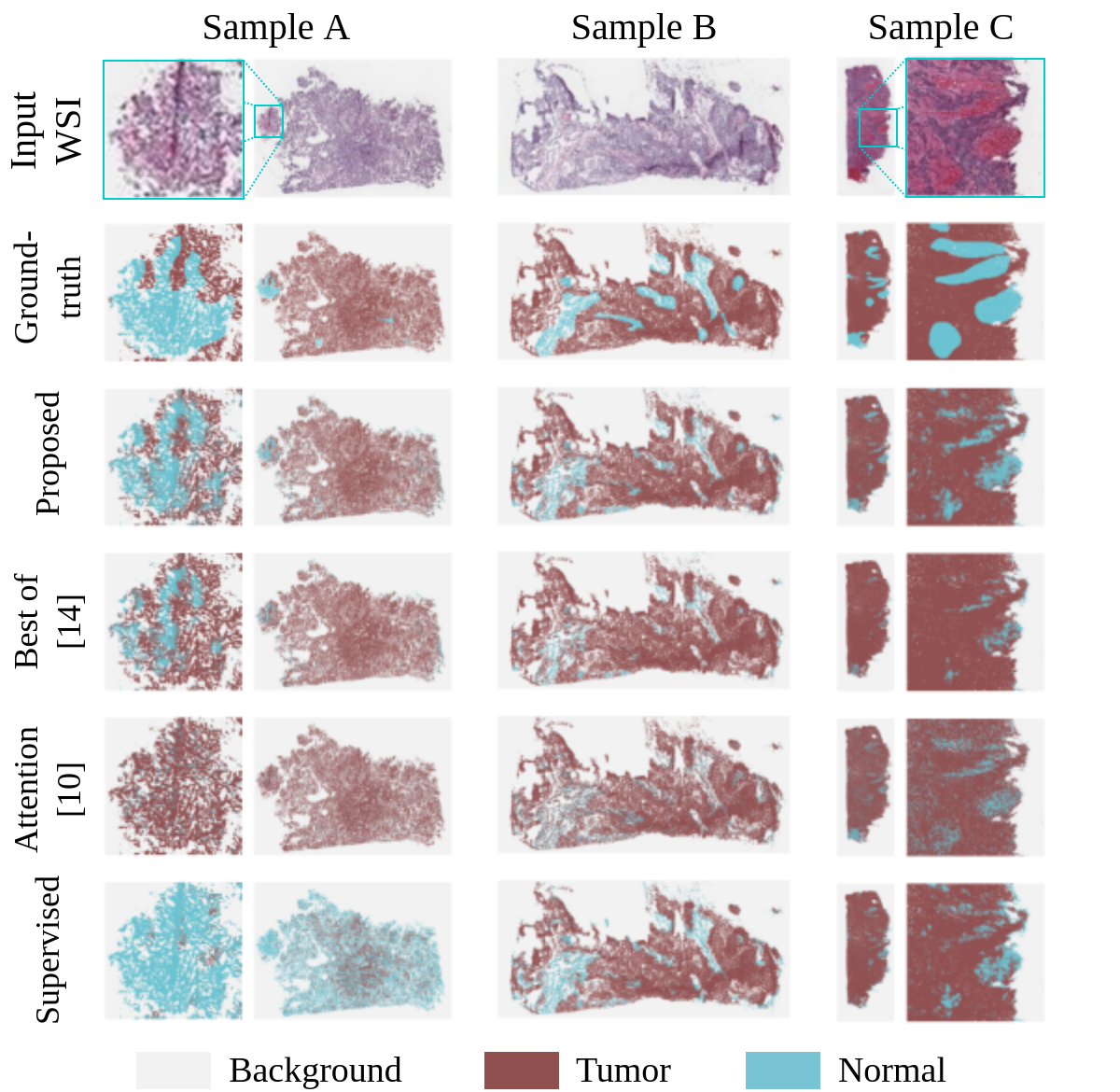}
\caption{Example of out-of-distribution segmentation maps for three testing whole slide images (Sample A, Sample B and Sample C).
The lines display, in that order, input WSIs, the pathologists curated ground-truth annotations, results from the proposed approach, results from the best performing configuration of~\cite{lerousseau2020weakly}, results from the attention-based MIL~\cite{ilse2018attention}, and results from the supervised approach.}
\label{fig:example_preds}
\end{figure}

\begin{table}[h]
\center
\begin{tabular}{lP{2.2cm}P{2.5cm}P{2.2cm}P{2cm}}
\hline
                       & Out-Of- & Out-Of- & \multicolumn{2}{c}{Out-Of-Domain}                                                        \\
Method & Distribution  & Location & DigestPath~\cite{li2019signet} & PAIP~\cite{kim2020paip} \\ \hline
AlphaBetaMIL \cite{lerousseau2020weakly}     &    &      &      &    \\
\quad ($\alpha=50, \beta=0$)  & 0.880  & 0.917 & 0.728 & 0.667\\
\quad ($\alpha=50, \beta=50$) &  0.847  & 0.876 & 0.562 & 0.612\\
\quad ($\alpha=75, \beta=0$) &  0.881  & 0.884 & 0.717 &  0.654\\
\quad Average             &  0.845  &  0.885 & 0.685 & 0.642\\
AttentionMIL~\cite{ilse2018attention}  & 0.892 & 0.864 & 0.540 & 0.545 \\
Supervised  & 0.905 & 0.886  & 0.534 & 0.620 \\
Proposed  &  \textbf{0.932} &  \textbf{0.946} & \textbf{0.779} & \textbf{0.671} \\ \hline
\end{tabular}
\caption{Performance of all benchmarked approaches on all datasets in area under the ROC curve. All results are computed on tissue pixels, \textit{i.e.} by completely discarding background pixels. \textbf{Bold} results highlight the best method for each testing cohort.}
\label{tab:results}
\end{table}

\subsection{Results}
\label{sec:results}
Testing results are reported in \autoref{tab:results}. Each model was inferred on all testing samples from both out-of-distribution, out-of-location, and out-of-domain testing sets. Then, each predicted whole slide image map was compared with the pathologists' curated ground-truth annotations, and AUC was computed on non-background pixels. For \cite{lerousseau2020weakly}, the 3 best performing configurations are displayed, with the "Average" representing the mean of all configurations; all results are displayed in supplementary Table 2.

The proposed approach achieved the best performance on all testing sets except for PAIP~\cite{kim2020paip} where it ranked second. 
Indeed, the error was reduced by more than $37\%$ compared to the best performing method that used binary annotations.
The proposed approach also appeared to better leverage percentage annotations with an error reduction higher than $28\%$ compared to the supervised approach that uses percentages.
Furthermore, the proposed approach obtained better generalization performances than the counterparts.
Indeed, the proposed approach is the top performer for all testing sets except for PAIP~\cite{kim2020paip} where it ranked second.
Notably, while AttentionMIL and the supervised approach loses performance in the out-of-location testing sets, the proposed approach achieved better than on holdout testing sets from tissue locations that were included in the training set.

We were interested in understanding the quality of the public percentages annotations that were used by the proposed approach. 
Statistical exploration revealed $44.9\%$ of non-$0\%$ annotations were multiple of $20\%$, whereas the ground-truth incidence should be closer to $5\%$, indicating that human annotators tending to round estimates to the closer multiple of $20\%$. Similarly, $89.1\%$ of annotations were found to be multiple of $5\%$ where incidence should be close to $20\%$. 
Moreover, pathologist inspection of the percentages annotations indicates that there seems to be confusion between tumor cells and tumor tissue. For instance, stromal tissue is often counted as tumor cells in annotations, whereas it can be considered as tumor tissue as a byproduct of tumor.
An histogram of percentages annotations is displayed in supplementary Figure 1 along with other distribution measures.
From this point of view, the proposed approach achieved remarkable performance although labels are noisy and bear systemic bias.

Some testing samples outputs are displayed in~\autoref{fig:example_preds}. Upon visual pathologists' qualitative inspection of the produced testing maps, most errors seemed to come from discriminating stromal from tumor cells, which are generally interlocked in tissues. Besides, less performing methods have more false positives on inflammatory cells which might reflect a bias in most locations, which is that immune cells are recruited in neoplastic tissues and are absent in normal tissues.

\section{Conclusion}
In this work, we leveraged weak annotations for training a tumor segmentation model for virtually all solid cancer subtypes. 
Experiments illustrate that the proposed approach performs better than traditional weakly supervised approaches, highlighting that the percentages annotations translate into performance gain.
Besides, the proposed approach seems robust with respect to bias and noise in percentages annotations and leads to tumor segmentation models of higher quality for performance and generalization.

Numerous extensions could be considered to further improve the performance of our versatile framework. 
First, performance could be enhanced by refining the input percentages annotations of the proposed approach and retraining with the same approach, in a transductive learning~\cite{vapnik1999overview} fashion.
In another perspective, although this work leveraged human annotations of percentages, the proposed approach could be used with input percentages computed by a model trained only with WSI binary annotations. The proposed approach would act as an annotation-free refining strategy for segmentation in WSI.
Finally, even though our work involves whole slide images, our generic approach could be used off-the-shelf for other modalities.

%
%
%

\newpage

\bibliographystyle{splncs04}
\bibliography{biblio}

\begin{thebibliography}{10}
\providecommand{\url}[1]{\texttt{#1}}
\providecommand{\urlprefix}{URL }
\providecommand{\doi}[1]{https://doi.org/#1}

\bibitem{campanella2019clinical}
Campanella, G., Hanna, M.G., Geneslaw, L., Miraflor, A., Silva, V.W.K., Busam,
  K.J., Brogi, E., Reuter, V.E., Klimstra, D.S., Fuchs, T.J.: Clinical-grade
  computational pathology using weakly supervised deep learning on whole slide
  images. Nature medicine  \textbf{25}(8),  1301--1309 (2019)

\bibitem{campanella2018terabyte}
Campanella, G., Silva, V.W.K., Fuchs, T.J.: Terabyte-scale deep multiple
  instance learning for classification and localization in pathology. arXiv
  preprint arXiv:1805.06983  (2018)

\bibitem{coudray2018classification}
Coudray, N., Ocampo, P.S., Sakellaropoulos, T., Narula, N., Snuderl, M.,
  Feny{\"o}, D., Moreira, A.L., Razavian, N., Tsirigos, A.: Classification and
  mutation prediction from non--small cell lung cancer histopathology images
  using deep learning. Nature medicine  \textbf{24}(10),  1559--1567 (2018)

\bibitem{deng2009imagenet}
Deng, J., Dong, W., Socher, R., Li, L.J., Li, K., Fei-Fei, L.: Imagenet: A
  large-scale hierarchical image database. In: 2009 IEEE conference on computer
  vision and pattern recognition. pp. 248--255. Ieee (2009)

\bibitem{dietterich1997solving}
Dietterich, T.G., Lathrop, R.H., Lozano-P{\'e}rez, T.: Solving the multiple
  instance problem with axis-parallel rectangles. Artificial intelligence
  \textbf{89}(1-2),  31--71 (1997)

\bibitem{foulds2010review}
Foulds, J.R., Frank, E.: A review of multi-instance learning assumptions
  (2010)

\bibitem{garcia2017review}
Garcia-Garcia, A., Orts-Escolano, S., Oprea, S., Villena-Martinez, V.,
  Garcia-Rodriguez, J.: A review on deep learning techniques applied to
  semantic segmentation. arXiv preprint arXiv:1704.06857  (2017)

\bibitem{he2016deep}
He, K., Zhang, X., Ren, S., Sun, J.: Deep residual learning for image
  recognition. In: Proceedings of the IEEE conference on computer vision and
  pattern recognition. pp. 770--778 (2016)

\bibitem{he2016identity}
He, K., Zhang, X., Ren, S., Sun, J.: Identity mappings in deep residual
  networks. In: European conference on computer vision. pp. 630--645. Springer
  (2016)

\bibitem{hou2016patch}
Hou, L., Samaras, D., Kurc, T.M., Gao, Y., Davis, J.E., Saltz, J.H.:
  Patch-based convolutional neural network for whole slide tissue image
  classification. In: Proceedings of the IEEE conference on computer vision and
  pattern recognition. pp. 2424--2433 (2016)

\bibitem{ilse2018attention}
Ilse, M., Tomczak, J.M., Welling, M.: Attention-based deep multiple instance
  learning. arXiv preprint arXiv:1802.04712  (2018)

\bibitem{kim2019deep}
Kim, R.H., Nomikou, S., Dawood, Z., Jour, G., Donnelly, D., Moran, U., Weber,
  J.S., Razavian, N., Snuderl, M., Shapiro, R., et~al.: A deep learning
  approach for rapid mutational screening in melanoma. bioRxiv p. 610311 (2019)

\bibitem{kim2020paip}
Kim, Y.J., Jang, H., Lee, K., Park, S., Min, S.G., Hong, C., Park, J.H., Lee,
  K., Kim, J., Hong, W., et~al.: Paip 2019: Liver cancer segmentation
  challenge. Medical Image Analysis  \textbf{67},  101854 (2020)

\bibitem{kingma2014adam}
Kingma, D.P., Ba, J.: Adam: A method for stochastic optimization. arXiv
  preprint arXiv:1412.6980  (2014)

\bibitem{lerousseau2020weakly}
Lerousseau, M., Vakalopoulou, M., Classe, M., Adam, J., Battistella, E.,
  Carr{\'e}, A., Estienne, T., Henry, T., Deutsch, E., Paragios, N.: Weakly
  supervised multiple instance learning histopathological tumor segmentation.
  In: International Conference on Medical Image Computing and Computer-Assisted
  Intervention. pp. 470--479. Springer (2020)

\bibitem{li2019signet}
Li, J., Yang, S., Huang, X., Da, Q., Yang, X., Hu, Z., Duan, Q., Wang, C., Li,
  H.: Signet ring cell detection with a semi-supervised learning framework. In:
  International Conference on Information Processing in Medical Imaging. pp.
  842--854. Springer (2019)

\bibitem{li2020deep}
Li, Z., Zhang, J., Tan, T., Teng, X., Sun, X., Li, Y., Liu, L., Xiao, Y., Lee,
  B., Li, Y., et~al.: Deep learning methods for lung cancer segmentation in
  whole-slide histopathology images--the acdc@ lunghp challenge 2019. arXiv
  preprint arXiv:2008.09352  (2020)

\bibitem{litjens20181399}
Litjens, G., Bandi, P., Ehteshami~Bejnordi, B., Geessink, O., Balkenhol, M.,
  Bult, P., Halilovic, A., Hermsen, M., van~de Loo, R., Vogels, R., et~al.:
  1399 h\&e-stained sentinel lymph node sections of breast cancer patients: the
  camelyon dataset. GigaScience  \textbf{7}(6),  giy065 (2018)

\bibitem{litjens2017survey}
Litjens, G., Kooi, T., Bejnordi, B.E., Setio, A.A.A., Ciompi, F., Ghafoorian,
  M., Van Der~Laak, J.A., Van~Ginneken, B., S{\'a}nchez, C.I.: A survey on deep
  learning in medical image analysis. Medical image analysis  \textbf{42},
  60--88 (2017)

\bibitem{martinez2007libvips}
Martinez, K., Cupitt, J.: Libvips: A fast image processing library with low
  memory needs  (2007)

\bibitem{metter2019trends}
Metter, D.M., Colgan, T.J., Leung, S.T., Timmons, C.F., Park, J.Y.: Trends in
  the us and canadian pathologist workforces from 2007 to 2017. JAMA network
  open  \textbf{2}(5),  e194337--e194337 (2019)

\bibitem{oquab2015object}
Oquab, M., Bottou, L., Laptev, I., Sivic, J.: Is object localization for
  free?-weakly-supervised learning with convolutional neural networks. In:
  Proceedings of the IEEE conference on computer vision and pattern
  recognition. pp. 685--694 (2015)

\bibitem{paszke2019pytorch}
Paszke, A., Gross, S., Massa, F., Lerer, A., Bradbury, J., Chanan, G., Killeen,
  T., Lin, Z., Gimelshein, N., Antiga, L., et~al.: Pytorch: An imperative
  style, high-performance deep learning library. arXiv preprint
  arXiv:1912.01703  (2019)

\bibitem{ronneberger2015u}
Ronneberger, O., Fischer, P., Brox, T.: U-net: Convolutional networks for
  biomedical image segmentation. In: International Conference on Medical image
  computing and computer-assisted intervention. pp. 234--241. Springer (2015)

\bibitem{schmauch2020deep}
Schmauch, B., Romagnoni, A., Pronier, E., Saillard, C., Maill{\'e}, P.,
  Calderaro, J., Kamoun, A., Sefta, M., Toldo, S., Zaslavskiy, M., et~al.: A
  deep learning model to predict rna-seq expression of tumours from whole slide
  images. Nature communications  \textbf{11}(1),  1--15 (2020)

\bibitem{stathonikos2013going}
Stathonikos, N., Veta, M., Huisman, A., van Diest, P.J.: Going fully digital:
  Perspective of a dutch academic pathology lab. Journal of pathology
  informatics  \textbf{4} (2013)

\bibitem{tomczak2015cancer}
Tomczak, K., Czerwi{\'n}ska, P., Wiznerowicz, M.: The cancer genome atlas
  (tcga): an immeasurable source of knowledge. Contemporary oncology
  \textbf{19}(1A), ~A68 (2015)

\bibitem{vapnik1999overview}
Vapnik, V.N.: An overview of statistical learning theory. IEEE transactions on
  neural networks  \textbf{10}(5),  988--999 (1999)

\bibitem{zarella2019practical}
Zarella, M.D., Bowman, D., Aeffner, F., Farahani, N., Xthona, A., Absar, S.F.,
  Parwani, A., Bui, M., Hartman, D.J.: A practical guide to whole slide
  imaging: a white paper from the digital pathology association. Archives of
  pathology \& laboratory medicine  \textbf{143}(2),  222--234 (2019)

\end{thebibliography}

%
%
%
%
%

\end{document}